\documentclass{aa}  
\usepackage{psfig}
\begin{document}
\thesaurus{ 1        
(09.02.1  
10.08.1   
12.04.2   
13.25.3  
)}
\title{The Cosmic X-ray Background spectrum observed with ROSAT and ASCA}
\author{T. Miyaji  \inst{1,2}
\and   Y. Ishisaki  \inst{3}
\and   Y. Ogasaka    \inst{4,5}
\and   Y. Ueda   \inst{5}
\and   M.J. Freyberg \inst{2}
\and   G. Hasinger \inst{1}
\and   Y. Tanaka  \inst{2}  
}

\offprints{T. Miyaji (tmiyaji@aip.de)}

\institute{
Astrophysikalisches Institut Potsdam, An der Sternwarte 16, D--14482
Potsdam, Germany
\and Max-Planck-Institut f\"ur extraterrestrische Physik,
D--85740 Garching b.\ M\"unchen, Germany
\and Tokyo Metropolitan University, 1-1 Minami-Osawa, Hachioji, 192-03, Japan
\and Code 662, NASA Goddard Space Flight Center, MD 20771, USA
\and The Institute of Space and Astronautical Sciences, 3-1-1, Yoshinodai, 
Sagamihara, 229, Japan  
}

\date{Received date; accepted date}

\titlerunning{Cosmic X-ray Background Spectrum} 
\authorrunning{Miyaji et al.}

\maketitle
\begin{abstract}
 We have made a series of joint spectral fits for two blank fields,
the Lockman Hole and the Lynx-3A field, where a significant amount of 
both {\it ASCA} and {\it ROSAT} PSPC data exist after thorough screenings. 
The {\it ASCA} SIS, GIS and {\it ROSAT} PSPC spectra from these fields 
have been  fitted simultaneously. Comparison at $E>1$ keV shows 
general agreement
within 10\% in the Lockman Hole data and a $20-30\%$ disagreement 
in the Lynx-3A data, indicating remaining observation-dependent 
systematic problems. 
In both cases, satisfactory fits have
been found for the overall 0.1-10 keV spectrum with an extragalactic
power-law component (or a broken power-law component with
steepening at $E<1$ keV), a hard thermal component with
plasma temperature of $kT^{\rm h}\approx0.14$ keV and a soft
thermal component  $kT^{\rm s}\approx0.07$ keV.

\end{abstract}
\section{Introduction}
\label{sec:intr}

 The global spectrum of the cosmic X-ray background 
is a primary piece of information for understanding its origin.  
The 3-50 keV CXRB spectrum observed with HEAO-1 A2 can be well 
described by a  $kT=40$ keV thin 
thermal plasma-like spectral shape (Marshall et al. \cite{marshall80}; 
Boldt \cite{boldt87}), which can be approximated by a power-law with 
a photon index of $\Gamma = 1.4$ in $E\la 10 keV$. 
{\it BBXRT} (Jahoda et al. \cite{jahoda92}) and {\it ASCA} 
(Gendreau et al. \cite{gend_spec}; Ishisaki et al. \cite{ishi98})
measurements show that this power-law component extends down 
to 1 keV, below which an excess is observed.  

 A number of authors report {\it ROSAT} measurements in
the 0.5-2 keV band (e.g. Hasinger \cite{has92}; 
Georgantopoulos et al. \cite{georg96}) 
and show about 30\% larger flux than the 
Gendreau et al.'s (\cite{gend_spec}) {\it ASCA} SIS 
result at 1 keV, with different slopes (see Hasinger \cite{has96} 
for review). The disagreement may be contributed by
the differences in the position/solid angle of the measured sky, 
problems arising from incomplete modelings, and/or calibration 
problems.

 In order to separate these effects, we have made a series 
of joint spectral fits of {\it ROSAT} PSPC, {\it ASCA} GIS, 
and {\it ASCA} SIS spectra from two fields of the sky, where 
sufficient amount of blank-sky data exist after thorough screening. 
Because of the limited data meeting the criteria, the work 
presented in this paper is not intended to determine the current 
best estimate of the global CXRB spectrum, but rather a comparison 
of measurements among {\it ASCA} and {\it ROSAT} instruments in 
the same parts of the sky with consistent modelings.  

 In Sect.\ref{sec:data}, we describe the {\it ASCA} and {\it ROSAT} 
data used in the analysis. Joint spectral fits 
are described in Sect.\ref{sec:fit}. The results are discussed
in Sect.\ref{sec:disc}.

\section{Observations}
\label{sec:data}

\subsection{Field Selection}

 The fields have been selected from blank sky fields with both {\it ASCA} 
and {\it ROSAT} observations. Severe screening criteria have been applied 
to the {\it ROSAT} data in order to minimize contamination 
from non-cosmic background.
This limits the number of usable fields for our purpose.  
The fields used for this work, with sufficient amount
of screened data, are the Lockman Hole (hereafter LH) and the Lynx-3A 
(hereafter LX) field.
 In order to use the same sky fields as much as possible for all
instruments, we have selected the PSPC and GIS data from the sky 
area approximately corresponding to the {\it ASCA} SIS FOV 
($22\arcmin \times 22\arcmin$).
The log of {\it ROSAT} and {\it ASCA} observations on these
fields are shown in Table \ref{tab:log}.  

\begin{table}[b]
\caption[ ]{Log of {\it ASCA} and {\it ROSAT} observations}
\begin{tabular}{lcc}
\hline\hline
      & Lockman Hole (LH) &  Lynx-3A (LX) \\ 
\hline
\multicolumn{2}{c}{{\it ASCA}  observations} \\
Pointing(s) $(\alpha, \delta)$ & (162.96,57.35) & (132.26,44.83)\\
(degrees,J2000) & (163.00,57.37)   &  \\
Dates obs. & 22-24 May 93 & 13-15 May 93 \\
Obs. Modes & \multicolumn{2}{c}{GIS:PH; SIS:4CCD Faint/Bright}\\
Exp. [s] (GIS/SIS) & 55352./47236. & 75476./68598.  \\
\multicolumn{2}{c}{{\it ROSAT} observations}\\
Seq.  & 900029p-4   & 900009p  \\  
Dates obs. & 26 Apr.-9 May 93 & 5-6 Apr. 91\\
Exp. [s] & 29588, 8181(PI$<$52)  & 23438 \\
Area [min$^2$] & 593.   &  520.  \\
\hline
\end{tabular}
\label{tab:log}
\end{table} 

\subsection{{\it ASCA} observation}

 The screenings of the {\it ASCA} data have been made with the 
following criteria: avoid SAA; earth elevation angle $>$ 5 deg 
(from night earth) or $>$25 deg (from day earth); avoid 2 minutes 
after the satellite day-night transition; magnetic cutoff rigidity 
$>$ 8 GeV/c. The events from hot and flickering pixels of SISs have
been removed. The data were taken when the GIS SP-discriminator was 
disabled, resulting in a high GIS background. Non-cosmic background 
(NXB) has been collected 
from night earth observations with the same magnetic cutoff 
rigidity screening, with 110,300 and 102,066 seconds of exposure 
for SIS and GIS respectively. The reproducibility of the
GIS NXB is about 3\% (Ishisaki \cite{ishisaki_t}) and the NXB constitutes
about 10\% and 30\% of the total count for SIS and GIS respectively.  
 
  We have created spectral response matrices using a ray-tracing 
assuming a uniform sky over a large field, taking into 
account the energy-dependent PSF and stray light problems of 
the {\it ASCA} instruments. For the GIS, Ishisaki (\cite{ishisaki_t}) 
estimates 
the systematic error of the absolute flux measurements using
this procedure at $\approx10\%$.
 
\subsection{{\it ROSAT} observation}

 In this work, we have used the time intervals with the 
sun-satellite-zenith angle greater than 120 deg 
(night time, avoiding contaminations by solar scattered X-rays),
the geographic latitude between -30 and 30 degrees (tropic), 
and the Master-Veto ($MV$) rate between 40 and 170 
(the particle background rate can be estimated using 
the Plucinsky et al. [\cite{plu_pbg}] model).
The model particle background (PBG) rates are about 5\% of 
the total events in channels 100$<$PI$<$200 (1-2 keV) 
and 1-2\% in 52$<$PI$<$99 (0.5-1 keV). Thus possible errors
associated with the particle background modeling
does not affect our analysis significantly.

 Light curves for the screened data have been further 
examined to exclude the time range with flux enhancements.
In particular, the light curve of low energy channels 
($E\leq 0.5$ keV or PI$\leq$51) for LH 
data show significant variation due to the Long Term 
Enhancements (LTE)(e.g. Snowden et al. \cite{snow94}), while no 
significant light curve variation has been detected for 
PI$\geq$51. Thus we have used the data from the period  
where the light curve is close to minimum for the LH 
PI$\leq$51 spectrum. Since the light curve minimum for
a particular pointed observation does not necessarily mean
no LTE contamination, limiting our analysis.
The EXSAS (Zimmermann et al. \cite{exsas}) package has been 
used for data screening, particle background subtraction, 
and spectrum preparation.   

\section{Joint spectral fits}
\label{sec:fit}

\begin{table*}[btp]
\caption[ ]{Results of the spectral fits (see text for model and 
parameter definitions)}
\begin{tabular}{lllc}
\hline\hline
Id. & Field & Best-fit parameters$^{\rm a}$ (with 90\% errors 
or an asterisk for fixed parameters) & $\chi^2/$ d.o.f\\
\hline
A1 & LH &  $\Gamma_{\rm P}$=1.5$\pm$0.3;$A_{\rm P}$=10.4$\pm$0.9;
          $\Gamma_{\rm G}$=1.49$\pm$0.05;$A_{\rm G}$=11.5$\pm$0.6;
          $\Gamma_{\rm S}$=1.38$\pm$0.04;$A_{\rm S}$=10.3$\pm$0.5 
          & 1.04 (258./248) \\

A2 & LH & $\Gamma$=1.43$\pm$0.04; $A_{\rm P}$=10.3$\pm$0.5;
                 $A_{\rm G}$=11.1$\pm$0.4; $A_{\rm S}$=9.9$\pm$0.4 
          & 1.06 (265./250)\\

A3 & LX &  $\Gamma_{\rm P}$=1.8$\pm$0.4;$A_{\rm P}$=11.9$\pm$1.1;
          $\Gamma_{\rm G}$=1.37$\pm$0.05;$A_{\rm G}$=8.6$\pm$0.4;
          $\Gamma_{\rm S}$=1.43$\pm$0.04;$A_{\rm S}$=8.1$\pm$0.3 
          & 1.08 (210./195)\\

A4 & LX & $\Gamma$=1.40$\pm$0.04; $A_{\rm P}$=10.9$\pm$0.6;
                 $A_{\rm G}$=8.9$\pm$0.4; $A_{\rm S}$=7.9$\pm$0.3 
          & 1.09 (215./197)\\

B1 & LH & $N_{\rm H20}=0.56*$;$\Gamma=1.42\pm.03$;$A_{\rm P}=10.0\pm.5$;
          $kT^{\rm h}=0.142^{+.007}_{-.004}$;
          $EI^{\rm h}_{\rm P}=18.8\pm1.3$;\\
   &    & $kT^{\rm s}=(5.7\pm.9)\times 10^{-2}$;
          $EI^{\rm s}_{\rm P}=9.0^{+2.5}_{-1.4}$;
          $N_{\rm G}=1.09\pm.05$;$N_{\rm S}=0.96\pm.04$
          & 0.97 (321./332)\\
B2 & LX & $N_{\rm H20}=2.75*$;$\Gamma=1.42\pm.03$;$A_{\rm P}=11.5\pm.7$;
          $kT^{\rm h}=.148\pm.005$;$EI^{\rm h}_{\rm P}=19.9\pm1.6$;\\
   &    & $kT^{\rm s}=(9.0\pm.8)\times 10^{-2}$;$EI^{\rm h}_{\rm P}=8.1\pm.3$;
          $N_{\rm G}=0.81\pm.04$;$N_{\rm S}=0.72\pm.03$ & 1.06 (351./331)\\
\hline
\end{tabular}
\label{tab:fits}

{$^{\rm a}$ Subscripts to parameters P:PSPC, G:GIS, S:SIS, none:joint 
for all instruments.} 
\end{table*} 
 
\subsection{Single power-law fits to $E>$ 1 keV data}

 We have used XSPEC for spectral analysis. 
The {\it BBXRT} and {\it ASCA} results (Sect. \ref{sec:intr}) 
show that the CXRB spectrum between 1-10 keV can be well-fitted
with a single power-law with $\Gamma\approx 1.4$. 
Thus we first made single power-law fits to the 
$E>1$ keV data as a simple check for consistency among
instruments. The PSPC, GIS and SIS pulse height spectra for 
$E>1$ keV (PSPC:1.0-2. keV, GIS:1.0-10. keV, SIS:1.0-7. keV) 
have been fitted with a model of the form $P(E)=AE^{-\Gamma}$, 
where $P(E)$ is the photon intensity in units of 
$[{\rm photons\,cm^{-2}s^{-1}keV^{-1}sr^{-1}}]$
and $E$ is the photon energy in keV. In the fit, the two parameters, 
$A$ (normalization) and $\Gamma$ (photon index) are varied separately
or only $A$ is separate while $\Gamma$ is joined. 
Fitting results are summarized in Table \ref{tab:fits} 
(fit id. A1--A4) with 90\% errors ($\Delta \chi^2=2.7$).
In A1--A4, all free parameters were 
varied during the error search. 
 
\begin{figure}[t]
\psfig{file=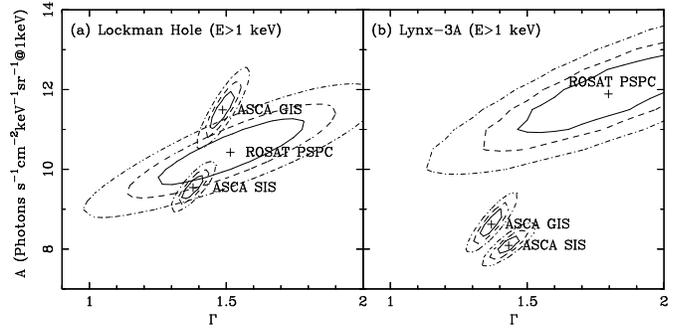,width=\hsize,angle=270}
\caption[]{The confidence contours for single power-law fits
to the $E>1$ keV data for the three instruments are shown 
for the LH (a) and LX (b) observations. The contours corresponding 
to $\Delta \chi^2=$ 2.3, 4.6, and 9.2 from the best-fit values
are plotted.}
\label{fig:cc}
\end{figure}

 Confidence contours in the $\Gamma-A$ space, considering
statistical errors only, for A1 and A3 are shown in 
Fig. \ref{fig:cc}. The disagreements among instruments in 
Fig. \ref{fig:cc} show the level of systematic errors. 
Possible sources of these systematic errors are discussed 
in Sect.\ref{sec:disc}. Adding an $E<1$ keV excess 
component to the model did not change the results 
significantly.
  
\subsection{Broad-Band fits (0.1 -- 10 keV)}

\begin{figure}[t]
\psfig{file=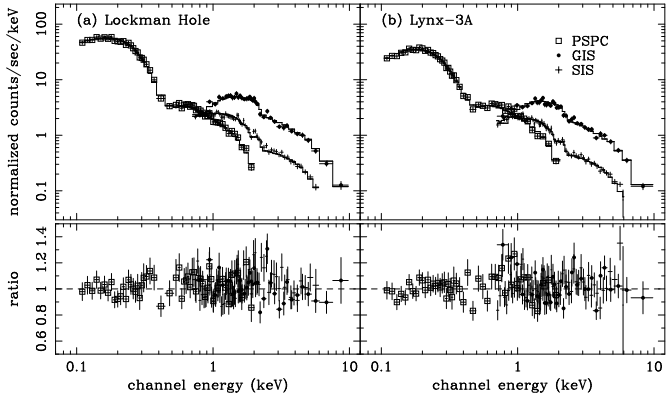,width=\hsize,angle=270}
\caption[]{The 0.1-10 keV pulse-height spectral data 
with $1\sigma$ errors, (appropriately rebinned for display), 
best-fit models, and the data/model ratio are 
shown for LH (a) (model B1) and LX (b) (B2). The data are 
marked according to the instrument as labeled.}
\label{fig:pha}
\end{figure}

 We have made joint fits to the overall {\it ROSAT -- ASCA} spectra 
over the 0.1 -- 10 keV range (0.1-2 keV with PSPC; 0.7-10 keV 
with GIS and 0.7-7 keV with SIS) considering the following
components: (1) an extragalactic power-law component with
parameters $A$ and $\Gamma$ (see above); (2) the hard thermal 
component, probably associated with the Galactic halo, 
with plasma temperature $kT^{\rm h}$[keV] and normalization $EI^{\rm h}$
in the XSPEC convention (per steradian); (3) the soft thermal component, 
from the Local Bubble with $kT^{\rm s}$ and $EI^{\rm s}$. 
Components (1) and (2) are absorbed by the interstellar gas with a 
hydrogen column density fixed to Galactic values 
($N_{\rm H20}=0.56 [LH]; 2.75 [LX]$ [$10^{20} {\rm cm^{-2}}$],
Dicky \& Lockman \cite{dicky90}). For components (2) and (3),
a Raymond \& Smith plasma (distributed as a part of XSPEC),
with the solar abundance was assumed. 
Spectral shape parameters and ratios of normalizations
of different spectral components have been joined for all
instruments. The overall normalizations have been
allowed to vary separately represented by a parameter $N_{\rm G}$ or 
$N_{\rm S}$, which is the relative normalization to the PSPC value. 
The pulse height spectra and models are shown in 
Fig. \ref{fig:pha} and fixed/best-fit parameters are listed in 
Table.\ref{tab:fits} (B1, B2). 
The 90\% error in  Table.\ref{tab:fits} are formal statistical errors, 
which have been derived with $\Gamma$ and temperatures  
fixed at nominal values while all the normalizations are fitted. 

 The good fits with the thermal components in the PSPC
$E<0.3$ keV data show that the after-pulse (AP) events 
can be neglected in these observations.
 Since the {\it ROSAT} spectra of resolved X-ray sources, mostly
extragalactic,  are steeper ($\Gamma\approx 2.0$, e.g. 
Hasinger et al. \cite{has_lnls}) and a larger flux than inferred from a
single power-law fit has already been resolved at $E\sim 0.5$ keV, we 
expect a turn up of the extragalactic component below $E=1$ keV, where 
the hard thermal component also start to emerge.
We could also obtain satisfactory fits with a model where the extragalactic
component has a break below 1 keV (fixed to $\Gamma=2.0$ 
at $E<1$ keV) with the main modification of the hard thermal
component normalization 
($\chi^2/\nu$=0.98 and 1.08 for LH and LX respectively). 
We could not find satisfactory fits with {\rm no hard thermal
component} but with an {\rm extragalactic component  excess below 1} keV
(either by a broken power-law or an addition of a steeper power-law
component).  This is because of a clear oxygen line feature 
around 0.5-0.6 keV in the PSPC spectrum 
(see also Hasinger \cite{has92}).

\section{Discussion}
\label{sec:disc}

 There is a bright variable source in LH with [1.2$\pm$.2] and [2.5$\pm$.7]
$\times 10^{-13} [{\rm erg\,cm^{-2}\,s^{-1}}]$ for the 0.7-2 keV and
2-7 keV bands respectively (Ogasaka \cite{oga_t}), consisting about 10\%
of the total ASCA fluxes in both bands. This source was much 
fainter during the PSPC observation. Thus
one should decrease the GIS and SIS normalizations about
$10\%$ lower when comparing with the PSPC data. In this
case, the agreement between PSPC and GIS is excellent and
falls well within statistical errors of each other. 
For both LH and LX, the SIS data consistently
show $\approx 10\%$ lower normalizations compared to GIS. This
might be caused by incomplete calibration for the radiation
damage of the SIS with the 4CCD mode, which can even
exist at this level after a few months after the launch,
when LH and LX were observed (Dotani et al. \cite{dotani95}).

 In the LX observation, a larger discrepancy exists.
The GIS and SIS normalizations are lower than the PSPC value
by  $\approx20\%$ and $\approx30\%$ respectively and slopes
are shallower. The ASCA
LX normalizations are also significantly lower than those
of LH.  There is no variable source which can cause this 
amount of discrepancy in LX.  The fact that the 0.1-10 keV
fit still show the disagreement of the normalization (see
$N_{\rm G}$ in B2) shows that this
is not a modeling problem (e.g. leak of the $E>1$ keV excess
with the PSPC energy resolution).  
One possible explanation is the LTE (e.g Snowden et al. \cite{snow94}), 
which is usually apparent in the $E<0.5$ keV channels of the PSPC data, 
but sometimes extends above 1 keV when the activity is high.  
There may also be an over-subtraction of the NXB background from 
the {\it ASCA} data. Furthermore, the instruments are not looking 
at exactly the same part of the sky. Due to the stray-light and 
PSF of the ASCA instruments, $\approx 40\%$ of the GIS/SIS flux
comes from outside of the designated FOV (estimated using our
ray-tracing program).  These effect may also contribute to this 
discrepancy.

\begin{figure}[t]
\psfig{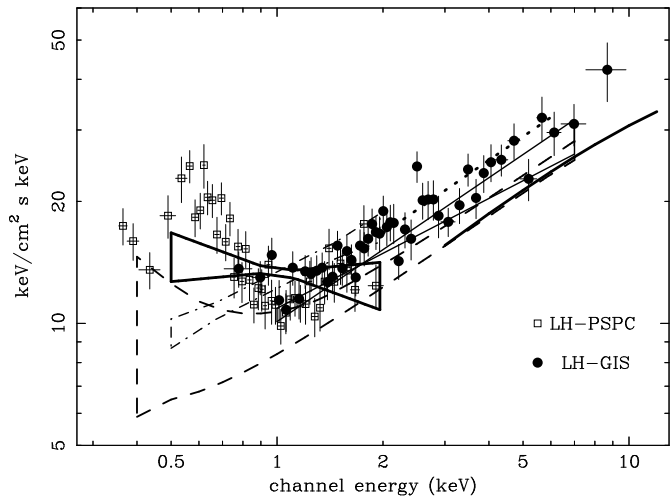}
\caption[]{The PSPC and GIS $E\,I(E)$ spectra (using a two 
power-law model as an appropriate smooth function for unfolding 
purpose) of LH are shown
and compared with previous measurements: the thick solid bowtie 
is from Hasinger (\cite{has92}); the 
dot-dashed bowtie from Georgantopoulos et al. (\cite{georg96}), both
used {\it ROSAT} PSPC. The dotted line is from
rocket measurements (McCammon \& Sanders \cite{maccamon90}). 
The long-dashed horn is from an {\it ASCA} SIS measurement by Gendreau
et al. \cite{gend_spec} and the thin solid bowtie is a joint {\it ROSAT} 
PSPC/{\it ASCA} SIS
analysis of QSF3 by Chen et al. (\cite{chen97}) for $E>1$ keV. 
The thick solid line represents the HEAO-1 A2 measurement by 
Marshall et al. (\cite{marshall80}).}
\label{fig:lo_comp}
\end{figure}

 The best consistency for a certain region of the sky from this work 
is seen for the PSPC and GIS data on LH. Thus it is instructive to 
compare the observed spectra of these with previous CXRB measurements. 
The comparison is  shown in Fig. \ref{fig:lo_comp}. 
  Fig. \ref{fig:lo_comp} shows a large excess at $E\sim 0.6$ keV
on the PSPC data, inconsistent with Gendreau et al.'s
{\it ASCA} SIS data. This may be partially due to the low Galactic 
column density of LH. The LH GIS data for $E\ga 2$ keV 
are above the HEAO-1 A2 and Gendreau et al. 
(\cite{gend_spec}) SIS values. 
We note, however, that about 10 \% of source fluctuation is expected over 
this small area. Since the
ASCA LH data contains a bright source ($\sim 5\times 10^{-13}
{\rm erg\,s^{-1}\,cm^{-2}}$ in 2-10 keV), this field should
be one of the brighter ones. We also note, however, that
an integration from the brightest source in the field to the
faintest source excluded in the collimator experiments
(e.g. $\approx2\times 10^{-11} {\rm erg\,s^{-1}\,cm^{-2}}$ in 2-10
keV for HEAO-1 A2 measurement by Marshall et al.) would
add $\approx10\%$ of intensity. A thorough treatment of source
fluctuations using a larger area and comparing spectra with
appropriate source removal will be presented in a future paper. 

 In summary, a close look at {\it ROSAT} and {\it ASCA} spectra
for the same regions of the sky have revealed systematic
errors caused by response calibration problems and non cosmic 
background subtraction of up to $\approx 20-30\%$ for one set of
observations. These probably caused
the reported disagreements between {\it ASCA} and {\it ROSAT} measurements
(Hasinger 1996), while modelings and sky selection can also 
contribute. We have obtained a fair description of the CXRB spectrum
over 0.1-10 keV range cosisting of a extragalactic power-law
component (either single or broken below 1 keV), hard and soft
thermal components with a satisfactory fit to all instruments. 
   
\acknowledgements
  TM was supported by a fellowship from the Max-Planck Society during 
his appointment at MPE. This work is partially supported by the 
Japanese-German collaboration on X-ray astronomy sponsored by JSPS 
and MPE. 


\begin{thebibliography}{}
\bibitem[1987]{boldt87} 
  Boldt E., 1987, Physics Reports 146, 215   

\bibitem[1997]{chen97} 
  Chen L.-W., Fabian A.C., Gendreau K.C., 1997, MNRAS 285, 499

\bibitem[1990]{dicky90} 
  Dicky J.M., Lockman F.J., 1990 ARAA 28, 215

\bibitem[1995]{dotani95} 
  Dotani T., Yamashita A., Ramussen A. et al.,  1995 
        ASCA News 3, 25

\bibitem[1995]{gend_spec} 
   Gendreau K.C., Mushotzky R.F., Fabian A.C. et al., 1995, 
        PASJ 47, L5 

\bibitem[1996]{georg96} 
  Georgantopoulos I., Stewart G., Shanks T., et al., 1996, 
        MNRAS 280, 276 

\bibitem[1992]{has92} 
   Hasinger G., 1992 in The X-ray Background, eds. X. Barcons, 
        A.C. Fabian, (Cambridge U. Press: Cambridge), 299

\bibitem[1993]{has_lnls} 
   Hasinger G., Burg R., Giacconi R., et al., 1993, A\&A 275, 1 

\bibitem[1996]{has96} 
   Hasinger G., 1996 A\&AS, 120, C607

\bibitem[1997]{ishisaki_t} 
   Ishisaki Y., 1997 Ph.D. thesis, University of Tokyo

\bibitem[1998]{ishi98} 
   Ishisaki Y. et al., 1998 ApJL, submitted

\bibitem[1992]{jahoda92} 
   Jahoda K. et al., 1992 in The X-ray Background, eds. X. Barcons, 
        A.C. Fabian, (Cambridge U. Press: Cambridge), 240

\bibitem[1990]{maccamon90} 
   McCammon D., Sanders W.T., 1990, ARAA 28, 657

\bibitem[1980]{marshall80} 
   Marshall F.E., Boldt E.A., Holt S.S. et al., 1980, ApJ 235, 4



\bibitem[1997]{oga_t} 
  Ogasaka, Y., 1997 Ph.D. thesis, Gakushuin University

\bibitem[1997]{plu_pbg} 
   Plucinsky P.P., Snowden S.L., Briel U.G., Hasinger G., \\
          Pfeffermann E.,  1993, ApJ 418, 519

\bibitem[1994]{snow94} 
   Snowden, S.L., McCammon D., Burrows D.N., Mendenhall, D.N., 
        1994 ApJ 424, 714  

\bibitem[1995]{exsas} 
    Zimmermann H.U., Becker W., Belloni T. et al., 1994, EXSAS 
          User's Guide, MPE Report 257, (Garching:MPE)
%
\end{thebibliography}
\end{document}